\documentclass[a4paper,11pt]{article}
\usepackage{jheppub} 
\usepackage{lineno}
\usepackage{xspace}
\usepackage{comment}
\usepackage{subfigure}

\newcommand{\jpsi} {\ensuremath{{\mathrm J}/\psi}}

\newcommand{\rhozero} {\ensuremath{\rho^0}\xspace}

\newcommand{\pp}           {pp\xspace}

\newcommand{\PbPb}         {\mbox{Pb--Pb}\xspace}

\newcommand{\pt}           {\ensuremath{p_{\rm T}}\xspace}

\newcommand{\y}         {\ensuremath{y}\xspace}

\newcommand{\nineH}        {$\sqrt{s}~=~0.9$~Te\kern-.1emV\xspace}
\newcommand{\seven}        {$\sqrt{s}~=~7$~Te\kern-.1emV\xspace}
\newcommand{\eight}        {$\sqrt{s}~=~8$~Te\kern-.1emV\xspace}
\newcommand{\twoH}         {$\sqrt{s}~=~0.2$~Te\kern-.1emV\xspace}
\newcommand{\twosevensix}  {$\sqrt{s}~=~2.76$~Te\kern-.1emV\xspace}
\newcommand{\five}         {$\sqrt{s}~=~5.02$~Te\kern-.1emV\xspace}
\newcommand{\fiveExactly}  {$\sqrt{s}~=~5$~Te\kern-.1emV\xspace}
\newcommand{\twosevensixnn}{$\sqrt{s_{\mathrm{NN}}}~=~2.76$~Te\kern-.1emV\xspace}
\newcommand{\fivenn}       {$\sqrt{s_{\mathrm{NN}}}~=~5.02$~Te\kern-.1emV\xspace}

\newcommand{\GeVc}         {Ge\kern-.1emV/$c$\xspace}
\newcommand{\MeVc}         {Me\kern-.1emV/$c$\xspace}
\newcommand{\TeV}          {Te\kern-.1emV\xspace}
\newcommand{\GeV}          {Ge\kern-.1emV\xspace}
\newcommand{\GeVtwo}       {Ge\kern-.1emV$^2$\xspace}
\newcommand{\MeV}          {Me\kern-.1emV\xspace}
\newcommand{\GeVmass}      {Ge\kern-.2emV/$c^2$\xspace}
\newcommand{\MeVmass}      {Me\kern-.2emV/$c^2$\xspace}

\arxivnumber{2410.06983} 

\title{\boldmath Machine learning opportunities for online and offline tagging of photo-induced and diffractive events in continuous readout experiments}

\author[1]{S. Ragoni\note{Corresponding author.}}
\author[1]{, J. Seger}
\author[1]{, C. Anson}
\author[1]{, D. Tlusty}
\affiliation{Creighton University,\\2500 California Plz, Omaha, \\NE 68178, United States, USA}

\emailAdd{simone.ragoni@cern.ch}
\emailAdd{jseger@creighton.edu}
\emailAdd{chrisanson5@gmail.com}
\emailAdd{tlusty@gmail.com}

\abstract{The increasing data rates in modern high-energy physics experiments such as ALICE at the LHC and the upcoming ePIC experiment at the Electron-Ion Collider (EIC) present significant challenges in real-time event selection and data storage. This paper explores the novel application of machine learning techniques, to enhance the identification of rare low-multiplicity events, such as ultraperipheral collisions (UPCs) and central exclusive diffractive processes. We focus on utilising machine learning models to perform early event classification, even before full event reconstruction, in continuous readout systems. We estimate data rates and disk space requirements for photoproduction and central exclusive diffractive processes in both ALICE and ePIC. We show that machine learning techniques can not only optimize data selection but also significantly reduce storage requirements in continuous readout environments, providing a scalable solution for the upcoming era of high-luminosity particle physics experiments.}

\begin{document}
\maketitle
\flushbottom

\section{Introduction}
\label{sec:intro}
The interest in photonuclear reactions is growing due to the efforts of the experiments at the CERN Large Hadron Collider (LHC) and at the Relativistic Heavy Ion Collider (RHIC). Results from the ALICE, ATLAS, CMS, LHCb and STAR collaborations show that these reactions can be studied within the context of ultraperipheral collisions (UPCs). In this type of events, the two projectiles from the colliding beams pass each other at impact parameters larger than the sum of the nuclear radii. As a consequence, hadronic interactions are highly suppressed, and the interactions are mediated through photon exchanges. 

The collaborations have already published results on vector meson photoproduction in UPCs, where the experimental signature is very clean: just the tracks of the particles from the decay of the vector meson, in an otherwise empty detector. A new venue for studies of UPCs consists in the study of inclusive UPCs, which instead feature particles only on one side of the detector, while on the side of the photon emitter the projectile remains intact.
Another interesting process is represented by central exclusive production (CEP) which can be purely diffractive. 

The relevant features of these processes are the very low rates (very small cross sections compared to purely hadronic counterparts for the current heavy ion experiments), and the very low multiplicities (typical multiplicities per unit of rapidity in central collisions at e.g. ALICE are of the order of $dN/d\eta \sim 2000$).   Because of these features, both the data rates and the typical event sizes are both quite small. In addition, these events have very clean topologies containing rapidity gaps.

Current and future particle physics experiments, i.e. the ALICE experiment and the ePIC experiment at the future EIC, are moving away from dedicated triggers and towards continuous readout. Since this leads to very large datasets, there is a need to optimise the disk space needed for the events without compromising the benefits brought by not having an online trigger. Machine learning algorithms are particularly suited for this purpose, since a machine can be trained to look for patterns, without enforcing a rigorous online or offline trigger.

This paper will present two practical examples of how shallow learning techniques can be employed for the online and offline tagging of photonuclear and diffractive events within the context of continuous readout.  The consequences for the data taking will be discussed. Finally, a comparison to current machine learning usage at particle physics experiments will be presented.

\section{Offline and online tagging using shallow learning}
\label{sec:online-offline}
While for ePIC photonuclear reactions will cover most of the data taking, in ALICE for example, photonuclear reactions happen alongside central \PbPb collisions, characterised by extreme multiplicities. This work will focus on classifying different categories of photonuclear reactions.

\begin{figure}[ht!]
	\begin{center}
		\subfigure[]{
			\label{fig:events-sl}
			\includegraphics[width=0.4\textwidth]{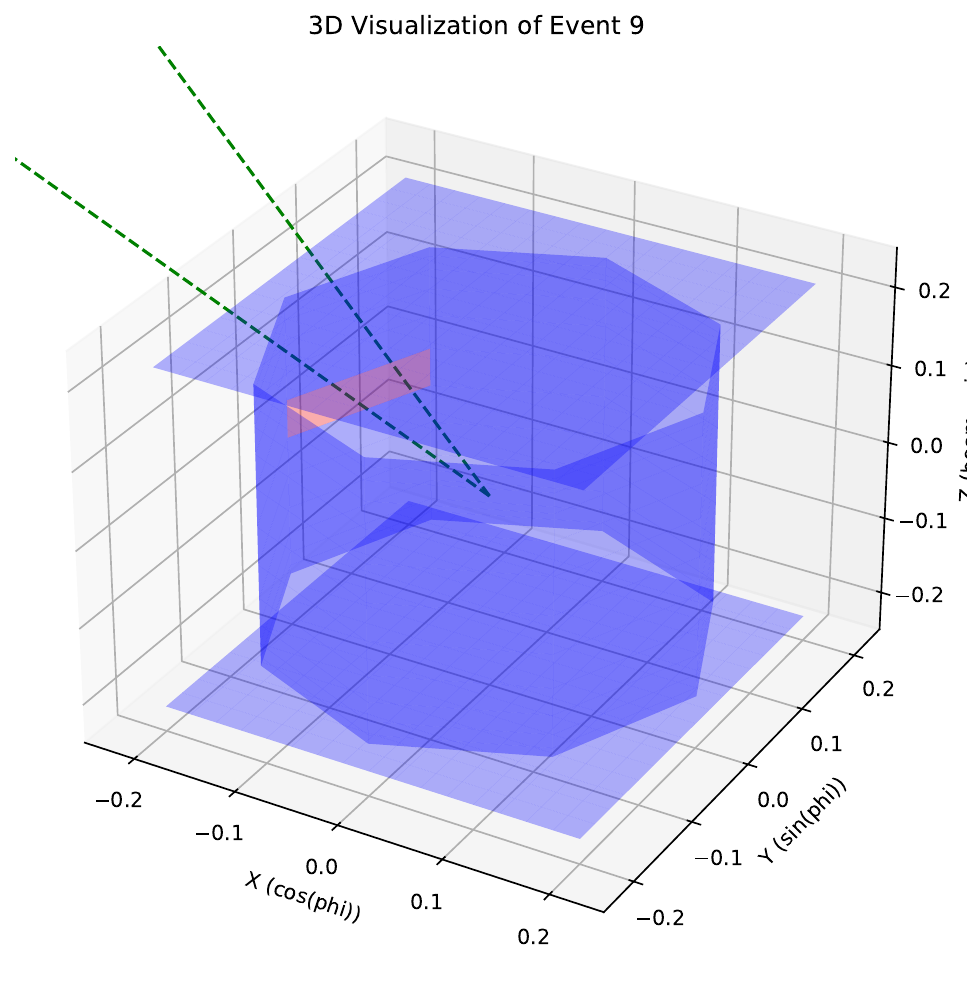}
		}
		\subfigure[]{
			\label{fig:events-dpm}
			\includegraphics[width=0.4\textwidth]{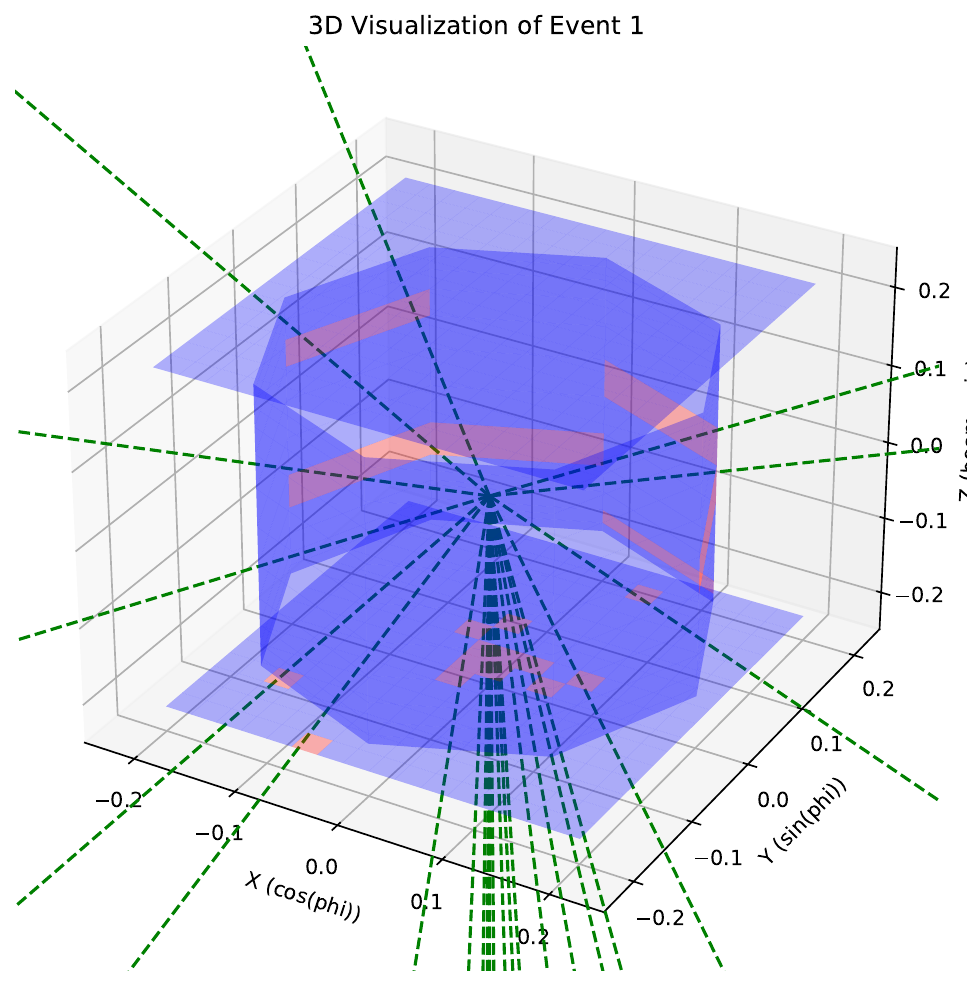}
		}\\		
	\end{center}
        \caption{Figure~\ref{fig:events-sl} and \ref{fig:events-dpm} show an event of coherent \jpsi\,\,production produced by STARlight and one inclusive produced by STARlight+DPMJET, respectively. The crossed detector modules are drawn in red, while the detector modules which were not crossed by the particle tracks are drawn in blue. }
	\label{fig:events}
\end{figure}

Photonuclear reactions may result in exclusive processes, such coherent vector meson production, which have been measured by the ALICE~\cite{alice-vector-meson}, CMS~\cite{cms-vector-meson}, LHCb~\cite{lhcb-vector-meson}, and STAR~\cite{star-vector-meson} collaborations, or inclusive UPC processes, such as inclusive vector meson photoproduction or photoproduction of $D^0$ and $\bar{D^0}$ pairs. The exclusive production of vector mesons can be simulated by STARlight~\cite{starlight} while inclusive events can be simulated by  STARlight+DPMJET~\cite{starlight-dpmjet} . To perform the studies presented in this paper, we have implemented the geometry of a toy detector composed of three parts, a single-layer central cylinder with a radius of 20 cm and a rapidity coverage of $|\eta| < 1.5$, which in the exercises here presented takes the role of a central tracker, and two planar detectors with a square geometry with a side length of 40 cm, which are positioned at 18 cm away from the interaction point, at the detector centre in this exercise, similar to the role of an endcap tracker. Both the cylinder and the planar detectors are segmented. The cylinder is divided in eight regions in azimuth $\phi$ and twelve  regions in pseudorapidity $\eta$, while the planar detectors are grids of sixteen by sixteen tiles. Figure~\ref{fig:events-sl} and \ref{fig:events-dpm} shows interaction of the particle tracks with the elements of the toy detector for an event of coherent \jpsi\,\,production produced by STARlight and one inclusive event produced by STARlight+DPMJET, respectively. 
In this section we offer two examples for the online and offline tagging of photonuclear and diffractive events.

\subsection{Classification before reconstruction}
\label{sec:before-reco}
The first example of shallow learning algorithms would identify photo-induced events during online data taking. The algorithm learns the differences in topology between possible processes.

In this exercise the information about the occupancy of each detector is gathered, which is akin to the data provided to a central trigger decision system, such as the one currently installed in ALICE, the Central Trigger Processor (CTP)~\cite{Kvapil:2021tuj}. The features that are available for the final decision are the following:
\begin{itemize}
    \item at least \textit{one} cylinder module was hit by a particle track;
    \item at least \textit{one} left planar module was hit by a particle track;
    \item at least \textit{one} right planar module was hit by a particle track;
    \item at least \textit{five} cylinder modules were hit by a particle track;
    \item at least \textit{five} left planar modules were hit by a particle track;
    \item at least \textit{five} right planar modules were hit by a particle track;
    \item at least 10\% of the cylinder modules were hit by a particle track;
    \item at least 10\% of the left planar modules were hit by a particle track;
    \item at least 10\% of the right planar modules were hit by a particle track.
\end{itemize}
These are all inputs that would be available in a very short time to the detector, and simple to implement.

The model we have trained for this purpose is a Random Forest classifier~\cite{random-forest} from the \textit{Scikit-learn} package~\cite{scikit-learn}. Such a classifier is suitable for deployment in particle physics experiments because it is very robust, easily traceable to physics selections, and not very prone to overfitting.

The training is performed over two categories of events.  One is a set of exclusive events produced by STARlight, containing a mixture of coherent \jpsi\,\,production and exclusive lepton pair production. The second category is a set of inclusive photonuclear events $A+A\rightarrow A+X$ generated by STARlight+DPMJET.

\begin{figure}[ht!]
	\begin{center}
		\subfigure[]{
			\label{fig:conf-before}
			\includegraphics[width=0.7\textwidth]{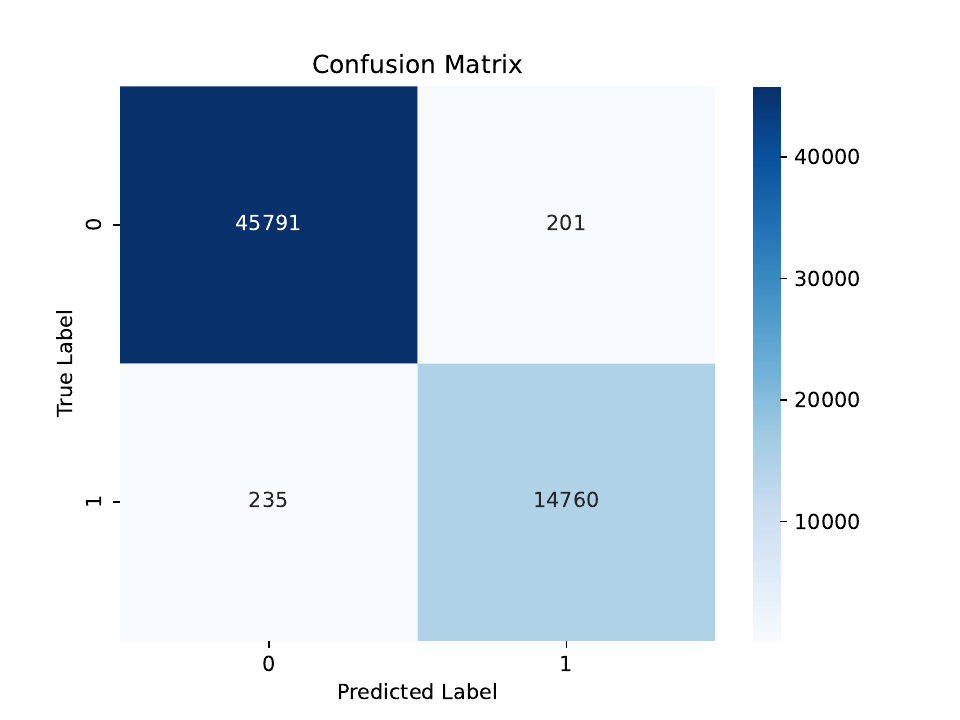}
		}\\
		\subfigure[]{
			\label{fig:roc-before}
			\includegraphics[width=0.7\textwidth]{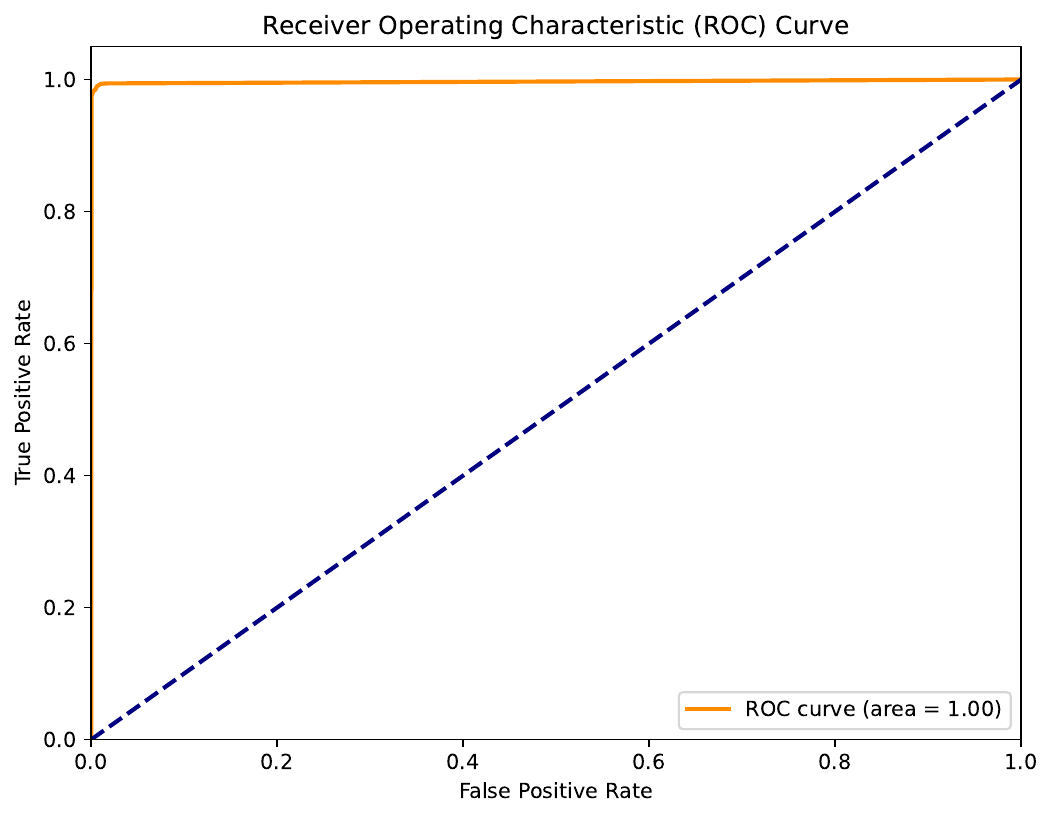}
		}\\		
	\end{center}
        \caption{Figure~\ref{fig:conf-before} and \ref{fig:roc-before} show the confusion matrix and the ROC curve of the Random Forest classifier applied to the detector occupancies, respectively. Very good performance is  achieved. }
	\label{fig:performance-before}
\end{figure}

We focus on these two categories because future EIC experiments are expected to feature essentially just these two processes. For LHC based experiments, such as ALICE, one might wish to also consider central and peripheral collisions (which however feature very high occupancy scenarios, and thus very different topologies from the two considered in the current exercise), 

The results of this model are shown in figure~\ref{fig:conf-before} and \ref{fig:roc-before} . The left panel shows the confusion matrix, which shows the performance of the model in classifying the exclusive and inclusive processes, and the right panel shows the ROC curves, which show the true-positive rate against the false-positive rate. Very good performance is achieved in this exercise, since the event topologies are very different in the cases under consideration.

\subsection{Classification after reconstruction}
\label{sec:after-reco}
This second example of applicability of shallow learning algorithms would identify and classify photo-induced events after reconstruction. The algorithm learns the differences in topology between possible processes.

In this exercise, the tracks are available after reconstruction. This means that the four-momenta of the particle tracks are available, and can be used as input to learning algorithms. As in the case of classification \textit{before} reconstruction, the model we have decided to train is a Random Forest classifier from the \textit{Scikit-learn} package from Python.  The training is performed over the same sets of exclusive events produced by STARlight and inclusive events produced by STARlight+DPMJET.

Since the four-momenta are all available, the features we have decided to use are the following:
\begin{itemize}
    \item the event multiplicity;
    \item the rapidity $\y$ gap on the left side of the toy interaction point;
    \item the $\y$ gap on the right side of the toy interaction point;
    \item the sum of all four-momenta of the reconstructed tracks;
    \item the transverse momentum \pt of the sum of all four-momenta of the reconstructed tracks;
    \item the mass of the sum of all four-momenta of the reconstructed tracks.
\end{itemize}

\begin{figure}[ht!]
	\begin{center}
		\subfigure[]{
			\label{fig:conf-after}
			\includegraphics[width=0.7\textwidth]{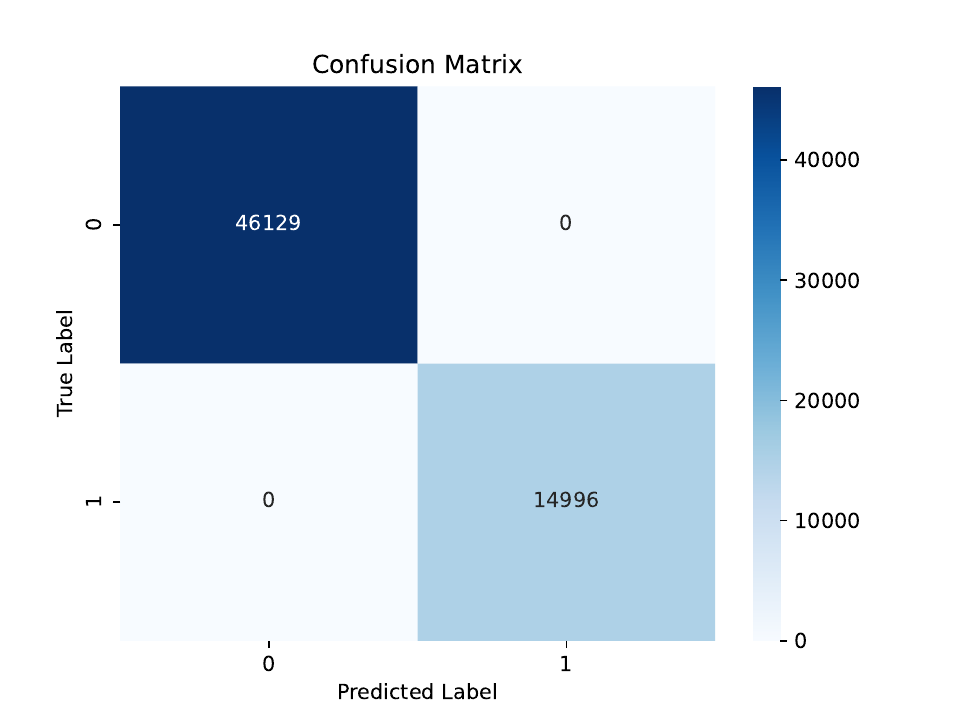}
		}\\
		\subfigure[]{
			\label{fig:roc-after}
			\includegraphics[width=0.7\textwidth]{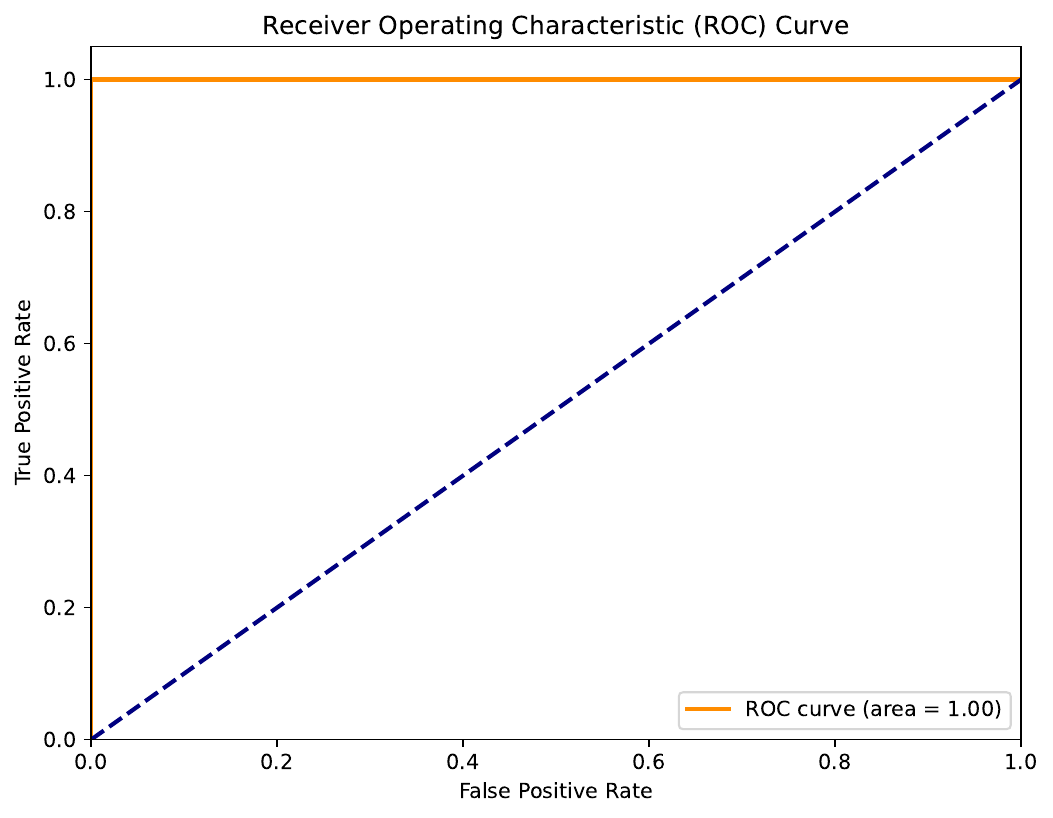}
		}\\		
	\end{center}
        \caption{Figure~\ref{fig:conf-after} and \ref{fig:roc-after} show the confusion matrix and the ROC curve of the Random Forest classifier applied to the tracks after reconstruction, respectively. Very good performance is  achieved. }
	\label{fig:performance-after}
\end{figure}

The results of this model are shown in figure~\ref{fig:conf-after} and \ref{fig:roc-after} for the confusion matrix and the ROC curves. Very good performance is again achieved in this exercise, due to the different event topologies.
The two Random Forest models applied to the cases \textit{before} and \textit{after reconstruction} have very high performances of about 100\%, since they operate in ideal conditions. The former shows how to perform this selection before reconstruction, e.g. through FPGA or software based algorithms, while the latter is purely software based, since it is after reconstruction. The two are then independent from each other, and could operate simultaneously within an experiment, with the former selecting online photoinduced events, and the latter further classifying it offline as exclusive or inclusive.

\section{Impact on the data taking}
\label{sec:data-taking}
While this work focuses on classifying different kinds of photonuclear reactions, i.e. the ePIC case, it can of course be extended quite naturally to also incorporate central \PbPb collisions, such as in ALICE, since they can be quite easily differentiated from photonuclear events in nature of their overwhelmingly higher multiplicities. 

The results of the previous classification exercises have important implications for the data-taking strategies in high-energy particle physics experiments. This is particularly relevant for continuous readout experiments, such as the ALICE experiment at the CERN LHC, and future experiments like the ePIC experiment at the Electron-Ion Collider (EIC).

In such experiments, there is a continuous stream of high-volume data originating from collision events. The ability to quickly classify and select interesting and rare events, such as diffractive and ultraperipheral collisions (UPCs) \cite{baltz-upc}, is crucial. These events are characterized by specific topologies, such as large rapidity gaps, which are not typically seen in more common, high-multiplicity collisions like those found in central heavy-ion events.

Being able to perform real-time classification of these low-multiplicity events, even before full reconstruction, presents a unique opportunity. It allows us to stream these events directly to either online or offline dedicated reconstruction systems. This is of particular importance because low-multiplicity events, such as those in diffractive and UPC processes, are more challenging to reconstruct in the high-multiplicity environments typical of heavy-ion experiments.
Real-time selection of these events would enable a faster analysis workflow, reducing the typical delays introduced by the computationally expensive reconstruction processes.

Currently in ALICE, data is stored and only afterward subjected to event skimming to identify interesting events. By contrast, the real-time application of classification models (such as Random Forest classifiers) during data taking offers a more dynamic and immediate method of event selection. This method could serve as a powerful tool in optimizing the software trigger, particularly for diffractive events, which are usually selected by vetoing signals in forward detectors. The application of machine learning techniques succeeds in an event selection which behaves in analogy to a very open online trigger, but with all possible combinations of trigger inputs - note also the lack of a need for stringent thresholds on forward rapidity detectors, usually applied as vetoes for exclusive and centrally diffractive events - thereby increasing the efficiency of event selection.

Furthermore, this real-time approach has the ability to reduce the data storage requirements. Given the nature of continuous readout experiments, the volume of data produced is immense, and the cost of storage is a limiting factor. Machine learning classifiers can be employed to preselect only the most relevant events for full storage and analysis, reducing the need to store every event in its entirety. This has the potential to downsize the overall data storage requirements for both current and future experiments.

With this in mind, we now turn to a detailed estimation of the data rates and storage requirements for both the ALICE experiment during LHC Run 3 and the ePIC experiment at the EIC. These estimates will help us understand the infrastructure and computing resources required for effective data processing and storage, particularly in light of the continuous data stream and the potential for real-time event selection.

\subsection{Data rates and storage estimates for ALICE}
The ALICE experiment is expected to process significantly higher data rates during LHC Run 3 due to upgraded detectors and improved data acquisition systems. Table~\ref{tab:alice_parameters} summarizes the key parameters for data taking in ALICE during Run 3 \cite{Rohr:2020qct}.
\begin{table}[h]
\centering
\begin{tabular}{|l|l|}
\hline
\textbf{Parameter} & \textbf{Value} \\ \hline
Raw data rate during \PbPb collisions & Up to 3.5~TB/s \\ \hline
Sustained data rate to storage after reduction & 90~GB/s \\ \hline
Run 3 duration (2022-2026) & 4 years for \PbPb \\ \hline
\PbPb data-taking weeks per year & 4 weeks for \PbPb \\ \hline
Total data expected for Run 3 & $\sim 100$~PB \\ \hline
\end{tabular}
\caption{Key parameters for ALICE during LHC Run 3,}
\label{tab:alice_parameters}
\end{table}

ALICE will operate approximately 4 weeks per year for Pb--Pb collision data taking. Using the sustained data storage rate of 90~GB/s over the four years of Run 3, the total data volume $V_{\text{Pb--Pb}}$ for the full Pb--Pb run is $\approx 870.9 \, \text{TB}$.

However, it can be expected that ALICE will record a total data volume of up to 100~PB of data during Run 3, taking into account additional data sources such as pp and p--Pb collisions, metadata, and backups. 

\subsubsection{Disk space for UPC and diffractive events}
Using a machine learning selection for UPC and central exclusive diffractive events will have a few benefits. Singling out these events will lead to a separate, immediate and not computationally expensive reconstruction, Also, in these types of events only a portion of the detector itself is involved, thus reducing the amount of raw data that needs to be stored, or even read out. This is particularly interesting, since not all low multiplicity events are actually of interest, i.e. UPC events or central exclusive diffractive events.

To estimate the disk space required for UPC and diffractive events, we scale the event rates relative to the total Pb--Pb cross section. Table~\ref{tab:cross_sections} lists the cross sections for different event types. In our estimation of UPC events, we take as a baseline the coherent photoproduction cross section of \rhozero mesons in ultraperipheral \PbPb collisions at \fivenn, which has been measured by the ALICE collaboration to be approximately 425 mb~\cite{ALICE:2020ugp}. To account for other photoproduction processes, we make a ballpark estimate by taking twice this cross section, leading to an estimated total cross section of approximately 850 mb. This estimate is based on the dominance of \rhozero production, which has a particularly large cross section compared to other photoproduction processes. Similar arguments can be raised for diffractive events in \pp, where cross sections are expected to be smaller, of the order of 1 to 10 mb \cite{CMS:2015inp}.

\begin{table}[h]
\centering
\begin{tabular}{|l|l|l|}
\hline
\textbf{Event Type} & \textbf{Cross Section ($\sigma$)} & \textbf{Comments} \\ \hline
Pb--Pb collisions & 8~barns & Baseline heavy-ion collisions \\ \hline
UPC events & 850 mb & Ballpark value (2x coherent \rhozero cross section) \\ \hline
Diffractive events & 1--10~mb & Single or double diffractive processes \\ \hline
\end{tabular}
\caption{Cross sections for Pb--Pb, UPC, and diffractive events at ALICE.}
\label{tab:cross_sections}
\end{table}

The event rates for UPC or diffractive events is proportional to their cross sections and the data storage rate can be scaled down further by considering the lower particle multiplicities in these events compared to central \PbPb collisions.

The data storage rate for these events can be estimated using the following equation, where $R_{\text{event}}$ represents the event rate for UPC or diffractive events:

\begin{equation}
R_{\text{event}} = R_{\text{Pb--Pb}} \times \left( \frac{\sigma_{\text{event}}}{\sigma_{\text{Pb--Pb}}} \right) \times \frac{1}{f_{\text{mult}}}
\end{equation}

Here, $f_{\text{mult}}$ represents an empirical multiplicity reduction factor which serves to scale down the data rates taking into account the very low multiplicities which are characteristics of photo-induced reactions and of purely diffractive events. We can estimate a  $f_{\text{mult, UPC}} \approx 10$--$100$ for UPC events and $f_{\text{mult, diff}} \approx 5$--$10$ for diffractive events. Using the ballpark estimate of UPC cross section from table~\ref{tab:cross_sections} and assume $f_{\text{mult}} = 10$--$100$, the data storage rate for UPC or diffractive events would fall between $95.6 \, \text{MB/s}$ and $956 \, \text{MB/s}$.

Therefore, a more accurate estimation is that UPC events contribute around 1\% of the total data volume. Thus, the data volume is estimated as:
\begin{equation}
V_{\text{UPC}} = \frac{1}{f_{\text{mult}}} \times\frac{850 \, \text{mb}}{8 \, \text{barns}} \times V_{\text{Pb--Pb}} \approx 1 \, \text{PB.} 
\end{equation}

In addition, by storing only the track information after reconstruction, the data volume can be further reduced. Assuming a data reduction factor of 10, the disk space requirements for UPC events are of the order of 0.1 PB. Thus, selecting online the interesting events would result in a significant reduction of the disk space requirements at the end of the chain, by reading out also only the involved detectors.

\subsection{Data rates and storage estimates for the ePIC experiment at the EIC}
The ePIC detector at the EIC will study electron-proton (e--p) and electron-ion (e--A) collisions. Table~\ref{tab:epic_parameters} outlines the key assumptions for the ePIC experiment \cite{Bernauer:2022moy}.
\begin{table}[h]
\centering
\begin{tabular}{|l|l|}
\hline
\textbf{Parameter} & \textbf{Value} \\ \hline
Raw data rate during e--p and e--A collisions & 100~GB/s \\ \hline
Data reduction factor after reconstruction & 10 \\ \hline
Total operating weeks per year & 30 weeks \\ \hline
Uptime for data collection & 50\% \\ \hline
Total data expected per year & $\sim 91 \text{ PB/y}$ \\ \hline
\end{tabular}
\caption{Key parameters for the ePIC experiment at the EIC.}
\label{tab:epic_parameters}
\end{table}

Assuming the EIC operates for 30 weeks per year with 50\% uptime, with a raw data rate of 100~GB/s and the assumed data reduction factor of 10, the yearly storage requirements for ePIC would be $V_{\text{EIC}} \approx  91 \, \text{PB/year}$.

\subsection{The impact of machine learning algorithms}
The strategies here presented allow for online and offline selections of photoinduced events. As such, they behave in a similar fashion to an online trigger and an offline skimmer.  The power of this technique consists of  enabling the capabilities of a very open trigger while selecting only the interesting events. In fact, the machine learning model will behave as the combination of all possible trigger inputs leading to the selection of that event, without being restricted to a particular configuration. With regard to the offline skimmer, the machine learning model will then learn how to best optimise the thresholds of the detector modules and all possible features known \textit{a posteriori} after reconstruction, being also able to limit the noise that traditional rectangular selections may not be able to avoid. 

To summarise the findings from above, machine learning may impact data collection in two ways. During online data taking, it may provide a significant tool to reduce data rates, since the diffractive, UPC, and single rapidity gap events can be directly streamlined to a dedicated reconstruction. Additionally, machine learning applied after event reconstruction can optimise data storage, and avoid saving data which are not strictly needed to validate the purity of the sample.

\subsection{Contemporary trigger developments in ATLAS and CMS}
Machine learning (ML) techniques are increasingly being integrated into the trigger systems of the CMS and ATLAS experiments to handle the high data volumes and complex event topologies in proton-proton collisions at the LHC. Recent Run-3 upgrades at the Level‑1 (L1) stage of both CMS and ATLAS have showcased the potential of machine learning under extreme latency constraints. However, almost all current implementations emphasize static spatial features such as calorimeter cluster shapes or operate in an unsupervised anomaly through the scoring regime.

At CMS, two flagship projects illustrate this trend. The AXOL1TL variational autoencoder 
processes full-event calorimeter images at 40~MHz collision rates and produces anomaly scores in fewer than 50~ns, enabling near-zero bias selection of rare signatures \cite{Duarte:2024lsg}. Meanwhile, the WOMBAT CNN targets boosted H$\rightarrow b\bar{b}$ signals by scanning tower-level energy deposits on Xilinx FPGAs
\cite{Bileska:2025jxv}. Both pipelines achieve remarkable purity and efficiency improvements but rely solely on spatial clustering information.

ATLAS’s Run 3 L1 calorimeter upgrade centers on the electromagnetic feature extractor (eFEX), which calculates multiple shower-shape and leakage variables from ten 25~ns-sampled energy readings per tower within 2.5~$\mu$s, yielding significant gains in electron/photon identification 
\cite{tmartin-2024}. Parallel firmware prototypes employ shallow autoencoders for trigger-level anomaly detection in low-multiplicity final states 
\cite{thong-2024}. 
The usage of anomaly detection for the total exclusive event is also proposed for anomaly detection at offline level e.g. at analysis stage, for UPC and diffractive events in \cite{Ragoni:2024ovv}.


The strategy here proposed goes instead beyond local anomaly detection. It enables the direct selection of fully exclusive event signatures in continuous readout environment or event at the L1 stage of triggered experiments. By making use of the physics signatures of low-multiplicity processes, like ultraperipheral collisions (UPCs) and diffractive events in ALICE and ePIC, it is possible to use e.g. event‐level occupancy and cluster metrics such as track multiplicities and rapidity gap sizes to train classifiers.  Such a classifier can distinguish rare exclusive processes (photoinduced, UPC, diffractive) from the high‐multiplicity background in continuous‐readout environments and enables the early identification of these low-multiplicity events during continuous readout, even before full event reconstruction. This contrasts with the traditional trigger-based approach in CMS and ATLAS, where events are selected primarily to reject backgrounds in high-multiplicity collisions. The ability to identify the entire event topology before full reconstruction in ALICE and ePIC presents a significant opportunity to also optimize storage by reducing the data volume for these low-multiplicity events.

\section{Conclusion}
This paper presents the first application of machine learning techniques in the context of photoinduced reactions, ultraperipheral collisions (UPCs), and diffractive events. While machine learning has been widely employed in high-energy physics experiments like CMS and ATLAS, particularly for identifying rare signals through spacial information, this is the first time these techniques have been explored to address the unique challenges posed by low-multiplicity processes.

The approach discussed here leverages lightweight classifiers to identify rare, low-multiplicity events in continuous readout environments, such as those present in the ALICE experiment at the LHC and the future ePIC experiment at the Electron-Ion Collider (EIC). By detecting distinctive event features, such as rapidity gaps and sparse track patterns, the proposed method provides a novel solution for real-time event selection.

Compared to CMS and ATLAS, which utilize machine learning in traditional trigger systems to handle dense event topologies, the approach described in this paper targets low-multiplicity processes also before full event reconstruction. This not only improves real-time decision-making but also offers significant reductions in data storage requirements, making it particularly well-suited for the continuous data streams in ALICE and ePIC.

In conclusion, this work opens a new avenue for using machine learning in the selection of rare processes, specifically for photoinduced reactions and diffractive events. The successful implementation of this technique could provide a scalable and efficient framework for handling the data challenges posed by future high-energy physics experiments.

\acknowledgments

This work was funded by the U.S. Department of Energy under contract number DE-FG02-96ER40991. This work has been presented as arXiv preprint \url{2410.06983}.

\subsection*{Data availability statement}
The data that support the findings of this study are openly available in GitHub at \url{https://github.com/siragoni/ML-upc}.

\bibliographystyle{JHEP}
\bibliography{biblio.bib}

\providecommand{\href}[2]{#2}\begingroup\raggedright\begin{thebibliography}{10}

\bibitem{alice-vector-meson}
{\scshape ALICE} collaboration, \emph{{Energy dependence of coherent photonuclear production of J/\ensuremath{\psi} mesons in ultra-peripheral Pb-Pb collisions at \fivenn}}, \href{https://doi.org/10.1007/JHEP10(2023)119}{\emph{JHEP} {\bfseries 10} (2023) 119} [\href{https://arxiv.org/abs/2305.19060}{{\ttfamily 2305.19060}}].

\bibitem{cms-vector-meson}
{\scshape CMS} collaboration, \emph{{Probing Small Bjorken-x Nuclear Gluonic Structure via Coherent J/\ensuremath{\psi} Photoproduction in Ultraperipheral Pb-Pb Collisions at \fivenn}}, \href{https://doi.org/10.1103/PhysRevLett.131.262301}{\emph{Phys. Rev. Lett.} {\bfseries 131} (2023) 262301} [\href{https://arxiv.org/abs/2303.16984}{{\ttfamily 2303.16984}}].

\bibitem{lhcb-vector-meson}
{\scshape LHCb} collaboration, \emph{{Study of exclusive photoproduction of charmonium in ultra-peripheral lead-lead collisions}}, \href{https://doi.org/10.1007/JHEP06(2023)146}{\emph{JHEP} {\bfseries 06} (2023) 146} [\href{https://arxiv.org/abs/2206.08221}{{\ttfamily 2206.08221}}].

\bibitem{star-vector-meson}
{\scshape STAR} collaboration, \emph{{Exclusive J/\ensuremath{\psi},~\ensuremath{\psi}(2s), and e+e\ensuremath{-} pair production in Au+Au ultraperipheral collisions at the BNL Relativistic Heavy Ion Collider}}, \href{https://doi.org/10.1103/PhysRevC.110.014911}{\emph{Phys. Rev. C} {\bfseries 110} (2024) 014911} [\href{https://arxiv.org/abs/2311.13632}{{\ttfamily 2311.13632}}].

\bibitem{starlight}
S.~Klein, J.~Nystrand, J.~Seger, Y.~Gorbunov and J.~Butterworth, \emph{\text{STARlight:} \text{A Monte Carlo} simulation program for ultra-peripheral collisions of relativistic ions}, \href{https://doi.org/10.1016/j.cpc.2016.10.016}{\emph{Computer Physics Communications} {\bfseries 212} (2017) 258}.

\bibitem{starlight-dpmjet}
O.~Djuvsland and J.~Nystrand, \emph{{Single and Double Photonuclear Excitations in Pb+Pb Collisions at $\sqrt{s_{NN}}=2.76$ TeV at the CERN Large Hadron Collider}}, \href{https://doi.org/10.1103/PhysRevC.83.041901}{\emph{Phys. Rev. C} {\bfseries 83} (2011) 041901} [\href{https://arxiv.org/abs/1011.4908}{{\ttfamily 1011.4908}}].

\bibitem{Kvapil:2021tuj}
J.~Kvapil et~al., \emph{{ALICE Central Trigger System for LHC Run 3}}, \href{https://doi.org/10.1051/epjconf/202125104022}{\emph{EPJ Web Conf.} {\bfseries 251} (2021) 04022} [\href{https://arxiv.org/abs/2106.08353}{{\ttfamily 2106.08353}}].

\bibitem{random-forest}
L.~Breiman, \emph{\text{Random Forests}}, \href{https://doi.org/10.1023/A:1010933404324}{\emph{Machine Learning} {\bfseries 45} (2001) 5}.

\bibitem{scikit-learn}
F.~Pedregosa, G.~Varoquaux, A.~Gramfort, V.~Michel, B.~Thirion, O.~Grisel et~al., \emph{\text{Scikit-learn:} \text{Machine Learning in Python}}, {\emph{Journal of Machine Learning Research} {\bfseries 12} (2011) 2825}.

\bibitem{baltz-upc}
A.J.~Baltz et~al., \emph{\text{The Physics of Ultraperipheral Collisions at the LHC}}, \href{https://doi.org/10.1016/j.physrep.2007.12.001}{\emph{Physics Reports} {\bfseries 458} (2008) 1}.

\bibitem{Rohr:2020qct}
{\scshape ALICE} collaboration, \emph{{Overview of online and offline reconstruction in ALICE for LHC Run 3}},  in \emph{{Connecting The Dots}}, 9, 2020 [\href{https://arxiv.org/abs/2009.07515}{{\ttfamily 2009.07515}}].

\bibitem{ALICE:2020ugp}
{\scshape ALICE} collaboration, \emph{{Coherent photoproduction of $\rho^{0}$ vector mesons in ultra-peripheral \PbPb collisions at $ \sqrt{{\mathrm{s}}_{\mathrm{NN}}} $ = 5.02 TeV}}, \href{https://doi.org/10.1007/JHEP06(2020)035}{\emph{JHEP} {\bfseries 06} (2020) 035} [\href{https://arxiv.org/abs/2002.10897}{{\ttfamily 2002.10897}}].

\bibitem{CMS:2015inp}
{\scshape CMS} collaboration, \emph{{Measurement of diffraction dissociation cross sections in pp collisions at $\sqrt{s}$ = 7 TeV}}, \href{https://doi.org/10.1103/PhysRevD.92.012003}{\emph{Phys. Rev. D} {\bfseries 92} (2015) 012003} [\href{https://arxiv.org/abs/1503.08689}{{\ttfamily 1503.08689}}].

\bibitem{Bernauer:2022moy}
J.C.~Bernauer et~al., \emph{{Scientific computing plan for the ECCE detector at the Electron Ion Collider}}, \href{https://doi.org/10.1016/j.nima.2022.167859}{\emph{Nucl. Instrum. Meth. A} {\bfseries 1047} (2023) 167859} [\href{https://arxiv.org/abs/2205.08607}{{\ttfamily 2205.08607}}].

\bibitem{Duarte:2024lsg}
J.M.~Duarte, \emph{{Novel machine learning applications at the LHC}}, \href{https://doi.org/10.22323/1.476.0012}{\emph{PoS} {\bfseries ICHEP2024} (2025) 012} [\href{https://arxiv.org/abs/2409.20413}{{\ttfamily 2409.20413}}].

\bibitem{Bileska:2025jxv}
M.~Bileska, \emph{{Design and FPGA Implementation of WOMBAT: A Deep Neural Network Level-1 Trigger System for Jet Substructure Identification and Boosted $H\rightarrow b\bar{b}$ Tagging at the CMS Experiment}},  other thesis, 5, 2025, [\href{https://arxiv.org/abs/2505.05532}{{\ttfamily 2505.05532}}].

\bibitem{tmartin-2024}
T.~Martin, \emph{\text{Triggering in ATLAS for Run 3}},  2024, \href{https://indico.stfc.ac.uk/event/1089/attachments/2292/4094/PPDseminar310724\_TimMartin.pdf}{https://indico.stfc.ac.uk/event/1089/attachments/2292/4094/PPDseminar310724\_TimMartin.pdf}.

\bibitem{thong-2024}
T.M.~Hong, \emph{\text{Anomaly detection with decision tree autoencoder}},  2024, \href{https://indico.global/event/805/contributions/23536/attachments/11118/16444/\\ TMHong20240513\_DPF\_autoencoder\_take2.pdf}{https://indico.global/event/805/contributions/23536/attachments/11118/16444/\\ TMHong20240513\_DPF\_autoencoder\_take2.pdf}.

\bibitem{Ragoni:2024ovv}
S.~Ragoni, J.~Seger and C.~Anson, \emph{{Zero-bias new particle searches using autoencoders in UPCs and diffractive events}},  \href{https://arxiv.org/abs/2411.00903}{{\ttfamily 2411.00903}}.

\end{thebibliography}\endgroup






\end{document}